\documentstyle[12pt,world_sci]{article}
\begin{document}

\title{{\bf NONCOMPACT GAUGE-INVARIANT LATTICE SIMULATIONS}}
\author{KEVIN CAHILL \thanks{Research supported
                by the U.~S. Department of Energy
                under contract DE-FG04-84ER40166;
                e-mail: kevin@cahill.phys.unm.edu.}
and GARY HERLING\thanks{Center for Advanced Studies,
e-mail: herling@bootes.unm.edu.}
\\
{\em Department of Physics and Astronomy,
University of New Mexico,\\
Albuquerque, New Mexico 87131-1156, U.~S.~A.}}

\maketitle
\setlength{\baselineskip}{2.6ex}

\begin{center}
\parbox{13.0cm}
{\begin{center} ABSTRACT \end{center}
{\small \hspace*{0.3cm}
We have applied a new gauge-invariant, noncompact,
Monte Carlo method to simulate
$U(1)$, $SU(2)$, and $SU(3)$ gauge theories
on $8^4$ and $12^4$ lattices.
The Creutz ratios of the Wilson loops
agree with the exact results for $U(1)$
for $\beta \ge 1$
apart from a renormalization of the charge.
The $SU(2)$ and $SU(3)$ Creutz ratios
robustly display quark confinement at $\beta = 0.5$
and $\beta = 1.5$, respectively.
At much weaker coupling, the $SU(2)$ and $SU(3)$ Creutz ratios
agree with perturbation theory after a renormalization
of the coupling constant.}}
\end{center}

\section{INTRODUCTION}
\par
The first gauge-invariant noncompact simulations
were carried out by Palumbo, Polikarpov, and Veselov~\cite{Palumbo92}
and were based on earlier work by Palumbo {\em et al.}~\cite{Palumboet}.
They saw a confinement signal~\cite{Palumbo92}.
Their action contains five terms,
constructed from two invariants,
and involves noncompact auxiliary fields and
an adjustable parameter.
\par
The present paper implements and tests
a new way of performing gauge-invariant
noncompact simulations.
This method is based upon a new
noncompact action that is exactly
invariant under lattice
gauge transformations~\cite{Cahillhep}.
The action is a natural discretization
of the classical Yang-Mills action
with auxiliary fields that are compact
group elements representing gauge transformations.
\par
We have used this method
to simulate $U(1)$, $SU(2)$, and $SU(3)$ gauge theories
on $8^4$ and $12^4$ lattices.
The Creutz ratios of Wilson loops
agree with the exact results for $U(1)$
for $\beta \ge 1$
apart from a renormalization of the charge.
The $SU(2)$ and $SU(3)$ Creutz ratios
clearly show quark confinement at $\beta = 0.5$
and $\beta = 1.5$, respectively.
At much weaker coupling, the $SU(2)$ and $SU(3)$ Creutz ratios
agree with perturbation theory with a renormalized
coupling constant.

\section{THE METHOD}
What constitutes a gauge transformation in this method?
To find out, we look at
the (massless) continuum fermionic action density
$i \bar \psi \gamma_\mu \partial_\mu \psi$\null.
A suitable discretization of this quantity is
$i \bar \psi (n) \gamma_\mu
[ \psi(n + e_\mu) - \psi(n)]/a$
in which $n$ is a four-vector
of integers representing an arbitrary vertex
of the lattice, $e_\mu$ is a unit vector
in the $\mu$th direction, and
$a$ is the lattice spacing.
The product of Fermi fields at the same
point is gauge invariant as it stands.
The other product of Fermi fields
becomes gauge invariant
if we insert a matrix $A_\mu(n)$ of gauge fields
\begin{equation}
{i \over a } \bar \psi (n) \gamma_\mu
[ (1 + i g a A_\mu(n)) \psi(n + e_\mu) - \psi(n)]
\end{equation}
that transforms under a gauge transformation
represented by the group elements
$U(n)$ and $U(n + e_\mu)$ in such a way that
\begin{equation}
1 + i a g A'_\mu(n) =
U(n) [ 1 + i a g A_\mu(n) ] U^{-1}(n + e_\mu).
\label{1+A'}
\end{equation}
The required behavior is
\begin{equation}
A'_\mu(n) = U(n) A_\mu(n) U^{-1}(n + e_\mu)
+ {i \over a g} U(n)
\left[ U^{-1}(n) - U^{-1}(n + e_\mu) \right].
\label{A'}
\end{equation}
\par
Let us define the lattice field
strength $F_{\mu\nu}(n)$ as
\begin{eqnarray}
F_{\mu\nu}(n) & = &
{ 1 \over a } [ A_\mu(n+e_\nu) - A_\mu(n) ]
- { 1 \over a } [ A_\nu(n+e_\mu) - A_\nu(n) ]
\nonumber \\
& & \mbox{} + i g  [ A_\nu(n) A_\mu(n+e_\nu)
- A_\mu(n) A_\nu(n+e_\mu) ]
\label{F}
\end{eqnarray}
which reduces to the continuum Yang-Mills
field strength in the limit $a \to 0$.
Under the gauge transformation (\ref{A'}),
this field strength transforms as
\begin{equation}
F'_{\mu\nu}(n) = U(n) F_{\mu\nu}(n) U^{-1}(n + e_\mu + e_\nu).
\label{F'}
\end{equation}
The field strength $F_{\mu\nu}(n)$
is antisymmetric
in the indices $\mu$ and $\nu$, but it is not
hermitian.
To make a positive plaquette action density,
we use the Hilbert-Schmidt norm of $F_{\mu\nu}(n)$
\begin{equation}
S = {1 \over 4 k} {\rm Tr} [F^\dagger_{\mu\nu}(n) F_{\mu\nu}(n)],
\label{S}
\end{equation}
in which it is assumed that the generators
of the gauge group have been orthonormalized as
${\rm Tr} ( T_a T_b ) = k \delta_{ab}$\null.
Because $F_{\mu\nu}(n)$ transforms
covariantly (\ref{F'}), this action density
is invariant under the
noncompact gauge transformation (\ref{A'}).
\par
The gauge transformation (\ref{A'})
with group element $U(n) = \exp(-i a g \omega^a T_a )$
maps the usual matrix of gauge fields
$A_\mu(n) = T_a A^a_\mu(n)$
into a matrix that generally lies outside the
Lie algebra of the gauge group,
although it does remain in the
algebra  to lowest (zeroth) order in the
lattice spacing $a$\null.
In our simulation
we use for the gauge fields
this more-general space of matrices.
We use the action (\ref{S})
in which the field strength (\ref{F})
is defined in terms of gauge-field
matrices that are of the form
\begin{equation}
A_\mu(n) = V A^0_\mu(n) W^{-1}
+ {i \over a g} V \left( V^{-1} - W^{-1} \right)
\label{newA'}
\end{equation}
where $ A^0_\mu(n) $ is a matrix of gauge fields
defined in the usual way,
$ A^0_\mu(n) \equiv T_a A^{a,0}_\mu(n) $.
Here the group elements $V$ and $W$
associated with the gauge field $A_\mu(n)$
are unrelated to those
associated with the neighboring
gauge fields $ A_\mu(n+e_\nu) $,
$ A_\nu(n) $, and $ A_\nu(n+e_\mu) $.
\par
We do not require the quantity
$ 1 + iga A_\mu(n) $ to be an element $ L_\mu(n) $
of the gauge group.
But if one did, then
the matrix $ A_\mu(n) $
of gauge fields would be
related to the link $ L_\mu(n) $ by
$A_\mu(n) =  ( L_\mu(n) - 1 )/( iga )$,
and the action (\ref{S})
defined in terms of the field strength
(\ref{F}) and this gauge-field matrix would be,
{\it mirabile dictu\/},
Wilson's action:
\begin{equation}
S =
{k - \Re \, {\rm Tr} L_\mu(n) L_\nu(n+e_\mu)
L^\dagger_\mu(n+e_\nu) L^\dagger_\nu(n)
\over 2 a^4 g^2k }.
\label{W}
\end{equation}

\section{RESULTS}
We have tested this method by applying it
to the $U(1)$, $SU(2)$, and $SU(3)$ gauge
theories on $8^4$ and $12^4$ lattices.
In our initial configurations,
the unitary matrices $V$ and $W$ were
set equal to the identity matrix
and the gauge fields $A^0_\mu$
were either zero or random.
We allowed at least 10,000 sweeps for
thermalization.
\par
For $U(1)$ and for $\beta \ge 1$,
our measured Creutz ratios
agreed with the exact ones
apart from finite-size effects
and a renormalization of the charge.
For instance at $\beta = 1$, we found
on the $8^4$ lattice
$\chi(2,2) =  0.142(1)$,
$\chi(2,3) =  0.098(1)$,
$\chi(3,3) =  0.047(1)$,
$\chi(2,4) =  0.085(1)$,
$\chi(3,4) =  0.030(1)$, and
$\chi(4,4) =  0.014(1)$\null.
The first three of these $\chi$'s are equal to the
exact Creutz ratios for a renormalized value of
$\beta_r = 0.93$; the last three are
smaller than the exact ratios for $\beta_r = 0.93$
due to finite-size effects by 6\%, 17\%, and 42\%, respectively.
\par
For $SU(2)$
on the $8^4$ lattice at $\beta = 0.5$,
we found
$\chi(2,2) =  0.835(3)$,
$\chi(2,3) =  0.85(1)$,
$\chi(3,3) =  0.9(2)$, and
$\chi(2,4) =  0.9(6)$
which robustly display confinement.
At $\beta = 1$, our six Creutz ratios
track those of tree-level perturbation
theory for a renormalized value of $\beta_r = 1.75$;
the finite-size effects are hidden by incipient
confinement.
\par
For $SU(3)$ at $\beta = 1.5$,
we found on the $8^4$ lattice
$\chi(2,2) =  1.175(3)$,
$\chi(2,3) =  1.16(2)$,
$\chi(3,3) =  1.7(16)$, and
$\chi(2,4) =  1.4(2)$,
and on the $12^4$ lattice
$\chi(2,2) =  1.171(4)$,
$\chi(2,3) =  1.14(3)$, and
$\chi(2,4) =  1.9(7)$\null.
At $\beta = 2$ we found on the $8^4$ lattice
$\chi(2,2) =  0.839(2)$,
$\chi(2,3) =  0.837(7)$,
$\chi(3,3) =  0.76(9)$, and
$\chi(2,4) =  0.87(3)$;
and on the $12^4$ lattice
$\chi(2,2) =  0.832(2)$,
$\chi(2,3) =  0.821(7)$,
$\chi(3,3) =  0.71(7)$, and
$\chi(2,4) =  0.80(2)$.
Within the limited statistics,
these results exhibit confinement.
At much weaker coupling, our ratios
agree with perturbation theory
apart from finite-size effects and
after a renormalization of the coupling constant.
\section*{ACKNOWLEDGMENTS}
We should like to thank
M.~Creutz,
G.~Marsaglia, F.~Palumbo,
W.~Press,
and K.~Webb
for useful conversations,
the Department of Energy for support under grant
DE-FG04-84ER40166, and B.~Dieterle for time
on his DEC Alpha.
\bibliographystyle{unsrt}

\end{document}